\documentclass[aps,12pt,pra,showpacs,preprint,a4paper]{revtex4}
\usepackage{amsmath}
\usepackage{graphicx}
\usepackage{subfig}
\usepackage{array}
\usepackage{amsfonts}
\usepackage{epsf}
\usepackage{epsfig}
\usepackage{amssymb}

\begin{document}
\title{ Analytical approximate bound state solution of Schr\"{o}dinger equation in $D$-dimensions with a new mixed class of potential for arbitrary $\ell$-state via asymptotic iteration method}
\author{\small Tapas Das}
\email[E-mail: ]{tapasd20@gmail.com}\affiliation{Kodalia Prasanna Banga High
 School (H.S), South 24 Parganas, 700146, India}
\begin{abstract}
The bound state solutions of the $D$-dimensional Schr\"{o}dinger equation for new mixed class of potential, $V(r)=\frac{V_1}{r^2}+\frac{V_2e^{-\alpha r}}{r}+V_3coth\alpha r+V_4\,,$ are studied within the framework of the Pekeris approximation for any arbitrary $\ell$-state. Asymptotic iteration method (AIM) is used for the work. The energy spectrum are obtained as well as their corresponding normalized eigenfunctions are derived in terms of generalized hypergeometric functions $\,_{2}F_{1}(a,b,c;z)$. It is shown that using the Pekeris approximation, present potential model is very much capable of deriving other well known potentials quite easily and corresponding solutions are in excellent agreement with the previous work carried out in literature.
\end{abstract}
\maketitle
\newpage
\section{I\lowercase{ntroduction}}
The exact solutions of the Schr\"{o}dinger wave equation are very important since they contain all the necessary information regarding the quantum system under consideration. Generally, we are always eager to search for a quantum system which is described by a confining potential. The confining potentials may have various form depending upon the interaction of the particles associated with the system. Unfortunately the list of confining potential is not lengthy in literature. The renowned confining potential systems are harmonic oscillator [1] and hydrogen like atom [2].\\
Over the past few decades interaction of particles are studied a lot, specially in high energy physics, chemical physics as well as atomic and nuclear physics. Such type of studies prescribe different potential models, that suit the nature of the interaction under investigation. The solutions of the Schr\"{o}dinger equation for these potentials help us to understand the quantum system very closely, specially when the relativistic effects are not important. Some of such works are listed in reference [3-10].\\
Recently, theoretical physicists have shown a great deal of interest in solving  $D$-dimensional  Schr\"{o}dinger equation for various spherically symmetric potentials [11-20]. These higher dimensional studies provide a general treatment of the problem in such a manner that one can obtain the required results in lower dimensions just dialing appropriate $D$. Many analytical as well as numerical techniques like the Laplace transform method [21-23], the Nikiforov-Uvarov method [24], the algebraic method [25], the $\frac{1}{N}$ expansion method [26], the path integral approach [27], the SUSYQM [28], the exact quantization rule [29] and others are applied to address Schr\"{o}dinger equation both for lower and higher dimensional cases. In addition to that, hyperbolic potentials, exponential-type potentials or their combinations have attracted a lot of interest of different authors [30-36], both for multidimensional and lower dimensional Schr\"{o}dinger equation. Bound state solutions of these potentials are very important in literature as they describe the different phenomenon like scattering, vibrational properties of molecules.\\
The worse fact is that, Schr\"{o}dinger equation is much difficult to solve analytically due to the presence of centrifugal term even for a simple potential that contains single exponential term such as Yukawa potential. The difficulty becomes much breath taking and mathematically clumsy when Schr\"{o}dinger equation is treated with hyperbolic type of potentials or combinations of them with exponential type of potentials. To ease out the scenario, approximation schemes like Pekeris are chosen [37-39]. Approximations are the only hope till some one finds any mathematical tool to solve the arbitrary $\ell$-state Schr\"{o}dinger equations exactly for hyperbolic or a mix of hyperbolic and exponential potentials. The good news is, using Pekeris approximation scheme for centrifugal term as well as for above said complicated potential terms, we can find bound state solutions to a very high degree of precision [40]. Using such approximation schemes various methods have been used by many researchers to find the approximate solutions of Schr\"{o}dinger equation for a physically accepted hyperbolic or mixed of hyperbolic and exponential potentials [41-42]. \\
Motivated by these type of works, in this paper another very practical method so called asymptotic iteration method (AIM) 
[43-45] has been employed  for special mixed type of potential 
\begin{eqnarray}
V(r)=\frac{V_1}{r^2}+\frac{V_2e^{-\alpha r}}{r}+V_3coth\alpha r+V_4\,,
\end{eqnarray}
to address the $D$-dimensional Schr\"{o}dinger equation. Here $V_i(i=1,2,3,4)$ are the potential parameters and $\alpha$ acts like the screening parameters. The interesting fact of this potential model is, under Pekeris approximation it can extract different notable potentials that are important in literature and studied by many authors.\\ So the purpose of this present work is to study the arbitrary $\ell$-state solutions of the $D$-dimensional Schr\"{o}dinger equation with the special mixed potential given by Eq.(1) using AIM within the frame work of Pekeris approximation scheme. The work is arranged as follows: Next section is for a brief out line of the AIM method. Section III is for the bound state energy spectrum and eigenstates. Section IV is devoted for the derivations of special cases where some well known potential models, their energy eigenvalues and eigenfunctions are derived. The conclusion of the present work comes at the section V.
\section{A\lowercase{symptotic} I\lowercase{teration} M\lowercase {ethod} (AIM)-B\lowercase{rief} O\lowercase{utline}}
AIM is proposed to solve the homogeneous linear second order differential equations of the form
\begin{eqnarray}
y_n^{''}(x)=\lambda_0(x)y_n^{'}(x)+s_0(x)y_n(x)\,,
\end{eqnarray}
where $\lambda_0(x)\neq 0$ and the prime denotes the derivative with respect to $x$, the extra parameter $n$ (integer) is thought as a radial quantum number (see section III). The variables, $\lambda_0(x)$ and $s_0(x)$ are sufficiently differentiable. To find a general solution to this equation, we differentiate Eq.(2) with respect to $x$ and find
\begin{eqnarray}
y_n^{'''}(x)=\lambda_1(x)y_n^{'}(x)+s_1(x)y_n(x)\,,
\end{eqnarray}
where
\begin{align}
\lambda_1(x)=\lambda_0^{'}(x)+s_0(x)+\lambda_{0}^2(x)\,, \nonumber\\
s_1(x)=s_0^{'}(x)+s_0(x)\lambda_0(x)\,.
\end{align}
Similarly, the second derivative of Eq.(2) provides
\begin{eqnarray}
y_n^{''''}(x)=\lambda_2(x)y_n^{'}(x)+s_2(x)y_n(x)\,,
\end{eqnarray}
where 
\begin{eqnarray}
\lambda_2(x)=\lambda_1^{'}(x)+s_1(x)+\lambda_0(x)\lambda_1(x)\,, \nonumber\\
s_2(x)=s_1^{'}(x)+s_0(x)\lambda_1(x)\,.
\end{eqnarray}
We can easily iterate Eq.(2) up to $(k+1)th$ and $(k+2)th$ derivatives, $k=1,2,3,....$\\
Therefore, we have
\begin{eqnarray}
y_n^{(k+1)}(x)=\lambda_{k-1}(x)y_n^{'}(x)+s_{k-1}(x)y_n(x)\,,\nonumber\\
y_n^{(k+2)}(x)=\lambda_k(x)y_n^{'}(x)+s_k(x)y_n(x)\,,
\end{eqnarray}
where 
\begin{eqnarray}
\lambda_k(x)=\lambda_{k-1}^{'}(x)+s_{k-1}(x)+\lambda_0(x)\lambda_{k-1}(x)\,,\nonumber\\
s_k(x)=s_{k-1}^{'}(x)+s_0(x)\lambda_{k-1}(x)\,,
\end{eqnarray}
which are called as the recurrence relation. 
Now from the ratio of the $(k+2)th$ and $(k+1)th$ derivatives, we have
\begin{eqnarray}
\frac{d}{dx}\ln[y_n^{(k+1)}(x)]=\frac{y_n^{(k+2)}(x)}{y_n^{(k+1)}(x)}=\frac{\lambda_k(x)[y_n^{'}(x)+\frac{s_k(x)}{\lambda_k(x)}y_n(x)]}{\lambda_{k-1}(x)[y_n^{'}(x)+\frac{s_{k-1}(x)}{\lambda_{k-1}(x)}y_n(x)]}\,.
\end{eqnarray}
For sufficiently large $k(>0)$, if
\begin{eqnarray}
\frac{s_k(x)}{\lambda_k(x)}=\frac{s_{k-1}(x)}{\lambda_{k-1}(x)}=\alpha(x)\,,
\end{eqnarray}
which is the ``asymptotic" aspect of the method, then Eq.(9) is reduced to
\begin{eqnarray}
\frac{d}{dx}\ln[y_n^{(k+1)}(x)]=\frac{\lambda_k(x)}{\lambda_{k-1}(x)}\,,
\end{eqnarray}
which yields
\begin{eqnarray}
y_n^{(k+1)}(x)=C_1\exp\left(\int{\frac{\lambda_k(x)}{\lambda_{k-1}(x)}} dx\right)=C_1\lambda_{k-1}(x)\exp\left(\int{[\alpha(x)+\lambda_0(x)]dx}\right)\,,
\end{eqnarray}
where $C_1$ is the integration constant and right hand side of the Eq.(12) is obtained by using Eq.(10) and Eq.(11). Inserting Eq.(12)
into Eq.(7), the first order differential equation is obtained as
\begin{eqnarray}
y_n^{'}(x)+\alpha(x)y_n(x)=C_1\exp\left(\int{[\alpha(x)+\lambda_0(x)]dx}\right)\,.
\end{eqnarray}
This is a first order differential equation which is very easy to solve and general solution of Eq.(2) can be obtained as:
\begin{eqnarray}
y_n(x)=\exp\left(-\int^x\alpha(x_1)dx_1\right)\left[C_2+C_1\int^x\exp\left(\int^{x_1}[\lambda_0(x_2)+2\alpha(x_2)]dx_2\right)dx_1\right]\,.
\end{eqnarray}
For a given potential, first the idea is to convert $D$-dimensional Schr\"{o}dinger equation to the form of Eq.(2) which gives $s_0(x)$ and $\lambda_0(x)$. Then, using the recurrence relations given by Eq.(8) parameters $s_k(x)$ and $\lambda_k(x)$ are obtained. The termination condition of the method in Eq.(10) can be arranged as
\begin{eqnarray}
\Delta_k(x)=\lambda_k(x)s_{k-1}(x)-\lambda_{k-1}(x)s_k(x)=0\,, \,\,\, k=1,2,3,\cdots
\end{eqnarray}
where $k$ is the iteration number. For the exactly solvable potential cases, using 
Eq.(15) the roots $\epsilon$ of $\Delta_k(x)$ are independent of $x$, and the vanishing of 
 $\Delta_k(x)$ gives us the exact analytical energy eigenvalues with radial quantum number $n$ is
equal to the iteration number $k$. For nontrivial potential cases that have no exact
solutions, $\Delta_k(x)$ depends on both $x$ and $\epsilon$. Then, equation $\Delta_k(x)=0$ is solved for a suitable chosen point $x = x_0$, the choice of which affects the convergence rate of the iteration. A suitable choice of $x$ gives the eigenvalue for small iteration numbers. If it is not chosen well, then the energy eigenvalues may be obtained at large iteration numbers $k$. A suitable point $x_0$ may be determined generally as the maximum value of the asymptotic eigenfunction or the minimum value of the potential [46] and the approximate energy eigenvalues are obtained from the roots of Eq.(15) for sufficiently great values of $k$ with iteration, 
for which $k$ is always greater than $n$ in these numerical solutions.\\
The general solution of Eq.(2) is given by Eq.(14). The first part of Eq.(14) gives the polynomial solutions that are convergent and physical, whereas the second part of Eq.(14) gives non-physical solutions that are divergent. Although Eq.(14) is the general solutions of Eq.(2), we take the coefficient of the second part $C_1=0$, in order to find the square integrable solutions. Therefore, the corresponding eigenfunctions can be derived from the following eigenfunction generator for exactly solvable potentials:
\begin{eqnarray}
y_n(x)=C_2\exp\left(-\int^x\frac{s_n(x_1)}{\lambda_n(x_1)}dx_1\right)\,,
\end{eqnarray}
where $n$ represents the radial quantum number.\\
Some comments are in order:
\begin{enumerate}
\item 
The terminating condition given by Eq.(10) is crucial for applying AIM for different problems. It is worth to mention here that, the above condition is imposed as an approximation or the iteration is assumed to terminate by the condition $\frac{s_k}{\lambda_k}=\frac{s_{k-1}}{\lambda_{k-1}}$. Without this condition we can not obtain the solution of the form of Eq.(14).
\item 
 Quantum mechanical eigenvalue problem for a particular potential model is governed by second order (Schr\"{o}dinger equation) differential equation without the first order derivative term. At a glance this seems $\lambda_0=0$ and AIM is little bit awkward for quantum mechanical eigenvalue type problem. The good news is, we can still generate the condition given by Eq.(10) if the eigenfunction or the solution is predetermined as $\psi(r)=\psi_0(r)f(r)$, where $\psi_0$ follows the well behaved boundary conditions of the problem and $f(r)$ follows the Eq.(2) with $x\equiv r$. Thus it is not hard to find $\lambda_0$, $s_0$ and hence the sequences $\lambda_k$, $s_k$ using Eq.(8). Now as we said earlier, if the problem is  exactly solvable, the terminating condition provides an expression that only depends on the eigenvalues $E$, that is to say independent of any space variable say $x$ (or $r$).\\
However, if the problem is not analytically solvable despite of the appropriate choice of $\psi_0(r)$, then the terminating condition produces for each iteration an expression that depends on both $E$ and concern space variable. In such a situation, the value of $x$ (or $r$) is so chosen that starting value of it stabilizes the process of iteration in such a manner that it doesn't oscillate but converges.
\item
The following theorem [47] is helpful to understand the method of AIM more better. {\bf Theorem:}
\textit{If the second order differential Eq.(2) has a polynomial solution of degree $n$, then $\lambda_n s_{n-1}-\lambda_{n-1}s_n=0$ }. $\lambda_n$ and $s_n$ are those obtain in Eq.(8) with $k\equiv n$. This could be a cross check of a problem, to justify the terminating condition well before using the asymptotic iteration method. In the above mentioned reference, all notable differential equations such as Hermite, Laguerre, and Bessel, Chebyshev, Legendre and others have been well examined and studied through the above theorem.  
\end{enumerate}
\section{S\lowercase{chr\"{o}dinger} e\lowercase{quation in} h\lowercase{yperspherical} c\lowercase{oordinates} - B\lowercase{ound} S\lowercase{tate} S\lowercase{pectrum}}
The $D$-dimensional time-independent Schr\"{o}dinger equation for a particle of mass $M$ with arbitrary angular momentum quantum number 
$\ell$ is given by [48] (in units of $\hbar=1$)
\begin{eqnarray}
\Big[\nabla_D^{2}+2M\Big(E_{n\ell}-V(r)\Big)\Big]\psi_{n\ell m}(r,\Omega_D)=0\,,
\end{eqnarray}
where $E_{n\ell}$ and  $V(r)$ denote the energy eigenvalues and potential. $\Omega_D$ within the argument of $n$-th state eigenfunctions $\psi$ denotes angular variables
$\theta_1,\theta_2,\theta_3,.....,\theta_{D-2},\varphi$. The Laplacian operator in hyperspherical coordinates is written as 
\begin{eqnarray}
\nabla_D^{2}=\frac{1}{r^{D-1}}\frac{\partial}{\partial r}(r^{D-1}\frac{\partial}{\partial r})-\frac{\Lambda_{D-1}^{2}}{r^2}\,,
\end{eqnarray}
where 
\begin{eqnarray}
\Lambda_{D-1}^{2}=-\Bigg[\sum_{k=1}^{D-2}\frac{1}{sin^2\theta_{k+1}sin^2\theta_{k+2}.....sin^2\theta_{D-2} sin^2\varphi}\times\left(\frac{1}{sin^{k-1}\theta_k}\frac{\partial}{\partial \theta_k}sin^{k-1}\theta_k\frac{\partial}{\partial \theta_k}\right)\nonumber\\+\frac{1}{sin^{D-2}\varphi}\frac{\partial}{\partial\varphi}sin^{D-2}\varphi\frac{\partial}{\partial\varphi}\Bigg]\,.
\end{eqnarray}
$\Lambda_{D-1}^{2}$ is known as hyperangular momentum operator.\\ The eigenvalues of $\Lambda_{D-1}^{2}$ are given by
\begin{eqnarray}
\Lambda_{D-1}^{2}Y_\ell^{m}(\Omega_D)=\ell(\ell+D-2)Y_\ell^{m}(\Omega_D)\,,
\end{eqnarray} 
where $Y_\ell^{m}(\Omega_D)$ is the hyperspherical harmonic.\\
By choosing a common ansatz for the eigenfunction in the form
\begin{eqnarray}
\psi_{n\ell m}(r,\Omega_D)=r^{-\frac{D-1}{2}}R_{n\ell}(r)Y_\ell^{m}(\Omega_D)\,,
\end{eqnarray}
Eq.(17) becomes
\begin{eqnarray}
\Bigg[\frac{d^2}{dr^2}+2M\Big(E_{n\ell}-V(r)\Big)-\frac{N_D^{\ell}}{r^2}\Bigg]R_{n\ell}(r)=0\,,
\end{eqnarray}
where $N_D^{\ell}=\frac{(D+2\ell-1)(D+2\ell-3)}{4}$.\\
Inserting the potential given by Eq.(1) into Eq.(22) leads to the following differential equation
\begin{eqnarray}
\Bigg[\frac{d^2}{dr^2}-2M(V_4-E_{n\ell})-\frac{2MV_1+N_D^{\ell}}{r^2}-2MV_2\frac{e^{-\alpha r}}{r}-2MV_3coth\alpha r\Bigg]R_{n\ell}(r)=0\,.
\end{eqnarray} 
It is obvious that Eq.(23), for $V_1\neq -\frac{N_D^{\ell}}{2M}$ and $V_2\neq 0$, does not have an exact solution due to the singular terms $1/r$ and $1/r^2$. Therefore, it is natural to look for a suitable approximating scheme. Here we will use the Pekeris approximation. In Pekeris approximation, during the study of hyperbolic and exponential potentials, generally the centrifugal term $1/r^2$ is replaced by $\frac{1}{r^2}\approx F(\alpha,r)$ with the condition $\alpha r<<1$. The function may have different choices, but the basic idea of deriving the unknown function $F(\alpha,r)$ comes from the expansion of the centrifugal term in a series of exponentials up to few suitable order depending on the intermolecular distance.
The renowned from of $F(\alpha,r)$ due to Pekeris is $\frac{\alpha^2}{sinh^2(\alpha r)}$. Nowadays, Slightly improved form of $F(\alpha,r)$ is taken as $\frac{\alpha^2}{sinh^2(\alpha r)}+\frac{\alpha^2}{3}$ [49]. We will not use the second choice because, the corrected term $\frac{\alpha^2}{3}$ does not create any big issue on the physics of the investigating problem.    
Exploring the $sinh(\alpha r)$ on the first choice, we have [38,39,42,50] 
\begin{subequations}
\begin{align}
\frac{1}{r^2}&\approx 4\alpha^2\frac{e^{-2\alpha r}}{(1-e^{-2\alpha r})^2}\,,\\
\frac{1}{r}&\approx 2\alpha\frac{e^{-\alpha r}}{(1-e^{-2\alpha r})}\,.
\end{align}
\end{subequations}
Now Eq.(23) changes to
\begin{eqnarray}
\Bigg[\frac{d^2}{dr^2}-2M(V_4-E_{n\ell})-(2MV_1+N_D^{\ell})\frac{4\alpha^2e^{-2\alpha r}}{(1-e^{-2\alpha r})^2}-4MV_2\alpha\frac{e^{-2\alpha r}}{(1-e^{-2\alpha r})}\nonumber\\-2MV_3\frac{(1+e^{-2\alpha r})}{(1-e^{-2\alpha r})}\Bigg]R_{n\ell}(r)=0\,.
\end{eqnarray}
Making a suitable change of variables as $s=e^{-2\alpha r}$, Eq.(25) takes the form with \\$R_{n\ell}(r)\Rightarrow R(s)$
\begin{eqnarray}
\frac{d^2R(s)}{ds^2}+\frac{1}{s}\frac{dR(s)}{ds}+\Bigg[-\frac{\epsilon_n^2}{s^2}-\frac{\gamma(\gamma-1)}{s(1-s)^2}-\frac{A}{s(1-s)}-\frac{B(1+s)}{s^2(1-s)}\Bigg]R(s)=0\,,
\end{eqnarray}
where the following abbreviations are used
\begin{subequations}
\begin{align}
\epsilon_n^2 &=\frac{M(V_4-E_{n\ell})}{2\alpha^2}\,,\\
\gamma(\gamma-1)&= 2MV_1+N_D^{\ell}\,, \\
A&=\frac{MV_2}{\alpha}\,,\\
B&=\frac{MV_3}{2\alpha^2}\,.
\end{align}
\end{subequations}
To solve Eq.(26) by using AIM, we assume the following physical eigenfunction satisfying the boundary conditions $R(s=0)=0$ and $R(s=1)=0$
\begin{eqnarray}
R(s)=s^c(1-s)^{\gamma}f_n(s)\,,
\end{eqnarray}  
where $c=\sqrt{\epsilon_{n}^2+B}$.\\
Inserting Eq.(28) into Eq.(26), we have the following linear second order homogeneous differential equation 
\begin{eqnarray}
\frac{d^2}{ds^2}f_n(s)=\left\{\frac{s(2c+2\gamma+1)-(2c+1)}{s(1-s)}\right\}\frac{d}{ds}f_n(s)+\left\{\frac{\gamma^2+2c\gamma+A+2B}{s(1-s)}\right\}f_n(s)\,.
\end{eqnarray}
Now it is easy to find the solution of Eq.(29) using AIM. Comparing with Eq.(2) we have
\begin{align}
\lambda_0(s)&=\frac{s\beta-\delta}{s(1-s)}\,, \nonumber\\
s_0(s)&=\frac{\eta}{s(1-s)}\,,
\end{align}
where
\begin{align}
\beta&=2c+2\gamma+1\,, &  \delta&=2c+1\,, &  \eta&=\gamma^2+2c\gamma+A+2B\,.
\end{align}
We may calculate $\lambda_k(s)$ and $s_k(s)$ from the recursion relation given by Eq.(8). This gives
\begin{eqnarray}
\lambda_1(s)=\lambda_{0}^{'}+s_0+\lambda_{0}^2=\frac{s^2(\beta^2+\beta-\eta)+s(\eta-2\delta-2\beta\delta)+\delta^2+\delta}{s^2(1-s)^2}\nonumber\\
s_1=s_{0}^{'}+s_0\lambda_0=\frac{s(2\eta+2\beta)-\eta(1+\delta)}{s^2(1-s)^2}........etc.
\end{eqnarray}
Here prime denotes the derivative with respect to $s$. Now Eq.(15) gives
\begin{eqnarray}
\Delta_1(s)=s_0(s)\lambda_1(s)-s_1(s)\lambda_0(s)=\frac{\eta(\eta+\beta)}{s^2(1-s)^2}\,.
\end{eqnarray}
The root of Eq.(33) gives the first value of $\epsilon_{k}^2$ i.e $\epsilon_{0}^2$. Similarly finding other $\Delta_k(s)$ we can explore different $\epsilon_{k}^2$. That means
\begin{eqnarray}
\Delta_1(s)=s_0(s)\lambda_1(s)-s_1(s)\lambda_0(s)=0\Rightarrow\epsilon_{0}^2=\frac{\gamma^4+2\gamma^2 A+(A+2B)^2}{4\gamma^2}\,, \nonumber\\
\Delta_2(s)=s_1(s)\lambda_2(s)-s_2(s)\lambda_1(s)=0\Rightarrow\epsilon_{1}^2=\frac{(\gamma+1)^4+2(\gamma+1)^2 A+(A+2B)^2}{4(\gamma+1)^2}\,,\nonumber\\
\Delta_3(s)=s_2(s)\lambda_3(s)-s_3(s)\lambda_2(s)=0\Rightarrow\epsilon_{2}^2=\frac{(\gamma+2)^4+2(\gamma+2)^2 A+(A+2B)^2}{4(\gamma+2)^2}\,,
\end{eqnarray}
.......and so on. Now using mathematical induction we can write the eigenvalues of the form
\begin{eqnarray}
\epsilon_{n}^2=\frac{(\gamma+n)^4+2(\gamma+n)^2 A+(A+2B)^2}{4(\gamma+n)^2}\,\,; n=0,1,2,3......
\end{eqnarray}
Now from the set of Eq.(27) the energy eigenvalues can be given as
\begin{eqnarray}
E_{n\ell}=V_4-\frac{1}{2M}\Bigg[\alpha^2(\gamma+n)^2+2MV_2\alpha+\frac{M^2(V_2+\frac{V_3}{\alpha})^2}{(\gamma+n)^2}\Bigg]\,.
\end{eqnarray}
It is to be noted that, by taking $s=z$ the Eq.(29) transforms to the differential equation
\begin{eqnarray}
z(1-z)\frac{d^2G(z)}{dz^2}+[u-z(v+w+1)]\frac{dG(z)}{dz}-vwG(z)=0\,,
\end{eqnarray}
where $v+w=2(c+\gamma)$ , $vw=\gamma^2+2c\gamma+A+2B=\eta$ , $u=2c+1$ and $G(z)\Rightarrow f_n(s)$.\\
Eq.(37) is satisfied by the hypergeometric function [51] $\,_{2}F_{1}(u,v,w;z)$.\\
Now we can derive the unnormalized eigenfunctions by using the generator given by Eq.(16)
\begin{eqnarray}
f_n(s)=(-1)^nC_2\frac{\Gamma(n+2c+1)}{\Gamma(2c+1)}\,_{2}F_{1}(-n,2(c+\gamma)+n,1+2c;s)\,,
\end{eqnarray}
where $\Gamma$ and $\,_{2}F_{1}$ are known as the gamma and the hypergeometric functions respectively[52].
Finally using Eq.(28) and Eq.(38) we can write the total unnormalized radial eigenfunctions as
\begin{eqnarray*}
R_{n\ell}(s)=K_{n\ell}s^c(1-s)^{\gamma}\,_{2}F_{1}(-n,2(c+\gamma)+n,1+2c;s)
\end{eqnarray*}
or
\begin{eqnarray}
R_{n\ell}(r)=K_{n\ell} e^{-2\alpha rc}(1-e^{-2\alpha r})^{\gamma}\,_{2}F_{1}(-n,2(c+\gamma)+n,1+2c;e^{-2\alpha r}) \,,
\end{eqnarray}
where $K_{n\ell}$ is the normalization constant. The complete eigenfunctions are given by Eq.(21) as
\begin{eqnarray}
\psi_{n\ell m}(r,\Omega_D)=K_{n\ell} r^{-\frac{D-1}{2}}e^{-2\alpha rc}(1-e^{-2\alpha r})^{\gamma}\,_{2}F_{1}(-n,2(c+\gamma)+n,1+2c;e^{-2\alpha r}) Y_\ell^{m}(\Omega_D)\,.
\end{eqnarray}
The normalization constant $K_{n\ell}$ can be evaluated from the condition 
$\int_0^\infty|R_{n\ell}(r)|^2dr=\frac{1}{2\alpha}\int_0^1|R_{n\ell}(s)|^2\frac{ds}{s}=1$. Now using the formula [51]
\begin{eqnarray}
\int_0^1 s^{2a-1}(1-s)^{2(b+1)}[\,_{2}F_{1}(-n,2(a+b+1)+n,1+2a; s)]^2ds\nonumber\\=\frac{(n+b+1)n!\Gamma(n+2b+2)\Gamma(2a)\Gamma(2a+1)}{(n+a+b+1)\Gamma(n+2a+1)\Gamma(n+2(a+b+1))}\,,
\end{eqnarray}
the normalization constant becomes
\begin{eqnarray}
K_{n\ell}=\Bigg[\frac{2\alpha(n+c+\gamma)\Gamma(n+2c+1)\Gamma(n+2(c+\gamma))}{n!(n+\gamma)\Gamma(n+2\gamma)\Gamma(2c)\Gamma(2c+1)}\Bigg]^{1/2}\,.
\end{eqnarray}
\section{D\lowercase{erivation} \lowercase{of} s\lowercase{pecial} c\lowercase{ases}}  
\begin{enumerate}
\item{\bf{ Yukawa Potential}}\\
In this case $V_1=V_3=V_4=0$ , $V_2\neq 0$ and $\alpha\neq 0$ i.e $V(r)=\frac{V_2e^{-\alpha r}}{r}$.\\
So Eq.(36) provides
\begin{eqnarray}
E_{n\ell}=-\frac{1}{2M}\Bigg[\alpha(\gamma+n)+\frac{MV_2}{\gamma+n}\Bigg]^2\,,
\end{eqnarray}
where $\gamma=\frac{D}{2}+\ell-\frac{1}{2}$.
To find the corresponding radial eigenfunctions here we have $B=0$, i.e $c=\epsilon_n=\frac{c_1}{2\alpha}$. This immediately gives the radial eigenfunctions from Eq.(39) as
\begin{eqnarray}
R_{n\ell}(r)=K_{n\ell} e^{-rc_1}(1-e^{-2\alpha r})^{\gamma}\,_{2}F_{1}(-n,2(\epsilon_n+\gamma)+n,1+2\epsilon_n;e^{-2\alpha r})\,, 
\end{eqnarray}
where $c_1=\sqrt{-2ME_{n\ell}}$.
This result is very much similar with the ref.[53].
As $\alpha\rightarrow 0$ the energy eigenvalues becomes as of Coulomb potential [21]
\begin{eqnarray}
E_{n\ell}=-\frac{MV_2^2}{2\Big(\frac{D}{2}+\ell-\frac{1}{2}+n\Big)^2}\,.
\end{eqnarray} 
Furthermore for $D=3$, the above equation gives the renowned bound state energy spectrum for Coulomb potential for Hydrogen like atom $E_{n\ell}=-\frac{MV_2^2}{2(n+\ell+1)^2}$ [48].
\item {\bf{ Mie-type Potential}}\\
In this case $V_3=0$ and $\alpha\rightarrow 0$ i.e $V(r)=\frac{V_1}{r^2}+\frac{V_2}{r}+V_4$. 
Using Eq.(36) the energy eigenvalues can be found as\\
\begin{eqnarray}
E_{n\ell}=V_4-\frac{MV_2^2}{2(\gamma+n)^2}\,,
\end{eqnarray}  
where $\gamma=\frac{1}{2}\Big[1+\sqrt{8MV_1+(D+2\ell-2)^2}\Big]$.\\
These results are in excellent agreement with the ref.[22]\\
The radial eigenfunctions in this situation become
\begin{eqnarray}
R_{n\ell}(r)=K_{n\ell} e^{-rc_2}(1-e^{-2\alpha r})^{\gamma}\,_{2}F_{1}(-n,2(\epsilon_n+\gamma)+n,1+2\epsilon_n;e^{-2\alpha r})\,, 
\end{eqnarray}
where $c=\epsilon_n=\frac{c_2}{2\alpha}$ and $c_2=\sqrt{2MV_4-2ME_{n\ell}}$.\\
The energy eigenvalues for {\bf{Kratzer-Fues potentials}} \Big($V(r)=\frac{V_1}{r^2}+\frac{V_2}{r}\Big)$ in $D$-dimensions can be found from Eq.(46) just setting $V_4=0$. Corresponding radial eigenfunctions are given by Eq.(47). These results also agree with the ref.[22] and ref.[44]. 
\item{\bf{ Manning-Rosen Potential}}\\
Taking $V_2=V_4=0$ and using the approximation given by Eq.(24a), we have the Manning-Rosen potential 
\begin{eqnarray}
V(r)=4V_1^{'}\frac{e^{-2\alpha r}}{(1-e^{-2\alpha r})^2}+V_3\frac{1+e^{-2\alpha r}}{1-e^{-2\alpha r}}\,,
\end{eqnarray}
where $V_1^{'}=V_1\alpha^2$.\\
The energy eigenvalues can be obtained from Eq.(36) in association with Eq.(27d) as
\begin{eqnarray}
E_{n\ell}=-\frac{\alpha^2}{2M}\Big[(\gamma+n)^2+\frac{4B^2}{(\gamma+n)^2}\Big]\,,
\end{eqnarray}
where $\gamma=\frac{1}{2}\Big[1+\sqrt{8M\frac{V_1^{'}}{\alpha^2}+(D+2\ell-2)^2}\Big]$. \\
Corresponding radial eigenfunctions are given by Eq.(39) as
\begin{eqnarray}
R_{n\ell}(r)=K_{n\ell} e^{-rc_3}(1-e^{-2\alpha r})^{\gamma}\,_{2}F_{1}(-n,2(c+\gamma)+n,1+2c;e^{-2\alpha r})\,, 
\end{eqnarray}
where $c=\sqrt{\epsilon_n^2+B}=\frac{c_3}{2\alpha}$ and $c_3=\sqrt{2MV_3-2ME_{n\ell}}$.\\
It is easy and straight forward to obtain the results of {\bf{Eckart-type potentials}} by setting $V_3=-V_{3}$. \\
The energy eigenvalues are same as Eq.(49) and the corresponding eigenfunctions are same as Eq.(50) except $c_3=\sqrt{-2MV_3-2ME_{n\ell}}$. These all results starting from Eq.(49) are consistent with the ref.[54]
\item{\bf{ Yukawa Plus Hulth\'{e}n Potential}}\\
Taking $V_1=0$,$V_3=-V_4=-\frac{V_0^{'}}{2}$ and $2\alpha\rightarrow b$ we have the Yukawa plus Hult\'{e}n potential 
\begin{eqnarray}
V(r)=\frac{V_2}{r}e^{-\frac{br}{2}}-V_0^{'}\frac{e^{-br}}{1-e^{-br}}\,.
\end{eqnarray}
The energy eigenvalues are given by
\begin{eqnarray}
E_{n\ell}=-\frac{1}{2M}\Bigg[\frac{b(\gamma+n)}{2}+\frac{M(bV_2-V_0^{'})}{b(\gamma+n)}\Bigg]^2 \,,
\end{eqnarray}
 where $\gamma=\frac{D}{2}+\ell-\frac{1}{2}$.\\
The eigenfunctions for this situation comes out from Eq.(39) as
\begin{eqnarray}
R_{n\ell}(r)=K_{n\ell} e^{-rc_4}(1-e^{-b r})^{\gamma}\,_{2}F_{1}(-n,2(c+\gamma)+n,1+2c;e^{-br})\,, 
\end{eqnarray}
where $c=\sqrt{\epsilon_n^2+B}=\frac{c_4}{b}$ and $c_4=\sqrt{-2ME_{n\ell}}$.\\
Now it is easy to obtain the results for {\bf{Hulth\'{e}n Potential}} i.e $V(r)=-V_0^{'}\frac{e^{-br}}{1-e^{-br}}$ \\by setting $V_2=0$ on the above calculations.\\
The energy eigenvalues for this case are
\begin{eqnarray}
E_{n\ell}=-\frac{1}{2M}\Bigg[\frac{b(\gamma+n)}{2}-\frac{MV_0^{'}}{b(\gamma+n)}\Bigg]^2 \,.
\end{eqnarray}
and the radial eigenfunctions are those obtained in Eq.(53). These results agree with the ref.[55] when $D=3$ 
\item{\bf{ Yukawa Plus Inverse Square Law Potential}}\\
In this case $V_3=V_4=0$ and the potential is of the form
\begin{eqnarray}
V(r)=\frac{V_1}{r^2}+\frac{V_2e^{-\alpha r}}{r}\,,
\end{eqnarray}
Eq.(36) helps to find the energy eigenvalues. Here it emerges as
\begin{eqnarray}
E_{n\ell}=-\frac{1}{2M}\Big[\alpha(\gamma+n)+\frac{MV_2}{\gamma+n}\Big]^2\,,
\end{eqnarray}
where $\gamma=\frac{1}{2}\Big[1+\sqrt{8MV_1+(D+2\ell-2)^2}\Big]$.\\
The corresponding radial eigenfunctions are 
\begin{eqnarray}
R_{n\ell}(r)=K_{n\ell} e^{- rc_5}(1-e^{-2\alpha r})^{\gamma}\,_{2}F_{1}(-n,2(\epsilon_n+\gamma)+n,1+2\epsilon_n;e^{-2\alpha r}) \,,
\end{eqnarray}
where $c=\epsilon_n=\frac{c_5}{2\alpha}$ and $c_5=\sqrt{-2ME_{n\ell}}$. All the above results are consistent with the ref.[56], if the approximate functions Eq.(24a-b) and potential parameters are  properly realized.
\item{\bf{ Quadratic Exponential-type Potential}}\\
Choosing $V_4=0, -2\alpha= \sigma$ and using the Pekeris approximation given by Eq.(24a) and Eq.(24b), the quadratic exponential-type potential is given by
\begin{eqnarray}
V(r)=\phi_0\frac{\xi_1e^{2\sigma r}+\xi_2e^{\sigma r}+\xi_3}{(e^{\sigma r}-1)^2}\,,
\end{eqnarray}
where $\xi_1, \xi_2, \xi_3$ are three adjustable parameters and $\phi_0$ is the depth of the potential. $\sigma$ acts like the range of the potential. The relations among them can be expressed as
\begin{subequations}
\begin{align}
\sigma V_2-V_3=\xi_1\phi_0\,,\\
\sigma^2V_1-\sigma V_2=\xi_2\phi_0\,,\\
V_3=\xi_3\phi_0\,.
\end{align}
\end{subequations}
Solving first two equations and using the third one, we have 
\begin{subequations}
\begin{align}
V_1=\frac{(\xi_1+\xi_2+\xi_3)\phi_0}{\sigma^2}\,,\\
V_2=\frac{(\xi_1+\xi_2)\phi_0}{\sigma}\,.
\end{align}
\end{subequations}
Eq.(36) provides the  energy eigenvalue equation as
\begin{eqnarray}
E_{n\ell}=\phi_0\xi_1-\frac{1}{2M}\Bigg[\frac{\sigma(\gamma+n)}{2}+\frac{M\phi_0(\xi_1-\xi_3)}{\sigma(\gamma+n)}\Bigg]^2\,,
\end{eqnarray} 
where $\gamma=\frac{1}{2}\Bigg[1+\sqrt{\frac{8M\phi_0(\xi_1+\xi_2+\xi_3)}{\sigma^2}+(D+2\ell-2)^2}\Bigg]$.\\
The radial eigenfunctions for quadratic exponential-type potential in $D$-dimensions become
\begin{eqnarray}
R_{n\ell}(r)=K_{n\ell}e^{-rc_6}(1-e^{\sigma r})^{\gamma}\,_{2}F_{1}(-n,2(c+\gamma)+n,1+2c;e^{\sigma r})\,,
\end{eqnarray}
where $c=-\frac{c_6}{\sigma}$ and $c_6=\sqrt{2M\xi_3\phi_0-2ME_{n\ell}}$. \\This results are consistent with the work listed in ref.[32]
\item{\bf{ Deng-Fun Potential}}\\
All the above results of quadratic exponential-type potential are useful in deriving the results of Deng-Fun Potential [57] which is expressed as
\begin{eqnarray}
V(r)=D_e\Big(1-\frac{\delta_0}{e^{\sigma r}-1}\Big)^2\,,
\end{eqnarray}
where $\delta_0=e^{\sigma r_e}-1$.\\ $D_e, \sigma, r_e$ are the dissociation energy, the range of the potential and the equilibrium inter nuclear distance receptively. Some author defines this potential as a  {\bf Generalized Morse Potential (GMP)} [48]\\
Eq.(58) provides Eq.(63) if we take $\phi_0=D_e, \xi_1=1, \xi_2=-2(1+\delta_0), \xi_3=(1+\delta_0)^2$. \\
Hence the energy eigenvalues for GMP come from Eq.(61) as
\begin{eqnarray}
E_{n\ell}=D_e-\frac{1}{2M}\Bigg[\frac{\sigma(\gamma+n)}{2}-\frac{MD_e\delta_0(2+\delta_0)}{\sigma(\gamma+n)}\Bigg]^2\,,
\end{eqnarray} 
where $\gamma=\frac{1}{2}\Big[1+\sqrt{\frac{8MD_e\delta_0^2}{\sigma^2}+(D+2\ell-2)^2}\Big]$.\\  
The radial eigenfunctions for generalized Morse potential in $D$-dimensions are given by
\begin{eqnarray}
R_{n\ell}(r)=K_{n\ell}e^{-rc_7}(1-e^{\sigma r})^{\gamma}\,_{2}F_{1}(-n,2(c+\gamma)+n,1+2c;e^{\sigma r})\,,
\end{eqnarray}
where $c=-\frac{c_7}{\sigma}$ and $c_7=\sqrt{2M(1+\delta_0)^2D_e-2ME_{n\ell}}$.
These results are excellent agreement with reference [48].
\end{enumerate}
\section{C\lowercase{onclusion}}
In this paper analytical approximate bound state solutions of the Schr\"{o}dinger equation  in $D$-dimensions has been studied for a special type of mixed potential with arbitrary $\ell$-state by means of the AIM and the approximation scheme of Pekeris to deal with the effective potential terms that contains $\frac{1}{r^2}$ and $\frac{1}{r}$. The eigenstates are represented by hypergeometric functions $\,_{2}F_{1}(a,b,c;z)$. Various well known potentials like, Yukawa, Mie-type, Manning-Rosen, Hulth\'{e}n, quadratic exponential type, Deng-Fun and others are extracted from the mixed potential model with their eigenfunctions and energy eigenvalues. The results are compared with the previous works and it is seen that Pekeris approximation scheme is quite capable of deriving results very near to the exact consequences.\\
The present potential model may become very customary to use in the field of quantum chemistry, molecular physics and particle interaction phenomenon of nuclear physics. It may be more interesting to study this potential in other methods also. It is obvious that the mathematical or graphical behaviors of present mixed class of potential, for suitable numerical values of $V_i(i=1,2,3,4)$ have not been examined in this paper. We expect future research to reveal this. We are also looking forward to the additional interesting topics associated with this new potential like scattering states, thermal or statistical studies, PT/non-PT aspects, just to name a few.
\section*{A\lowercase{cknowledgments}}
T.Das dedicates this work to his mother for her love and care. The author is grateful to the kind referees for the positive suggestions and critics which have greatly improved the present paper. 

\end{document}